\documentstyle[aps,prl,12pt,epsf]{revtex}

\begin{document}
\baselineskip 24pt

\renewcommand{\thetable}{\Roman{table}}

\normalsize
\begin{flushright}
FERMILAB-PUB-99/301-E\\
\end{flushright}
\Large
\begin{center}
\begin{bf}
Production of $\Upsilon(1S)$ Mesons from $\chi_b$ Decays in $p \overline{p}$
Collisions at $\sqrt{s} = 1.8$ TeV
\end{bf}
\end{center}
\normalsize
%
\font\eightit=cmti8
\def\r#1{\ignorespaces $^{#1}$}
\hfilneg
\begin{sloppypar}
\noindent
T.~Affolder,\r {21} H.~Akimoto,\r {43}
A.~Akopian,\r {36} M.~G.~Albrow,\r {10} P.~Amaral,\r 7 S.~R.~Amendolia,\r {32} 
D.~Amidei,\r {24} J.~Antos,\r 1 
G.~Apollinari,\r {36} T.~Arisawa,\r {43}
% ...
\end{sloppypar}
\vskip .026in
\begin{center}
(CDF Collaboration)
\end{center}

\vskip .026in
\begin{center}
\r 1  {\eightit Institute of Physics, Academia Sinica, Taipei, Taiwan 11529, 
Republic of China} \\
\r 2  {\eightit Argonne National Laboratory, Argonne, Illinois 60439} \\
\r 3  {\eightit Istituto Nazionale di Fisica Nucleare, University of Bologna,
I-40127 Bologna, Italy} \\
\r 4  {\eightit Brandeis University, Waltham, Massachusetts 02254} \\
\r 5  {\eightit University of California at Los Angeles, Los 
Angeles, California  90024} \\  
\r 6  {\eightit Instituto de Fisica de Cantabria, University of Cantabria, 
39005 Santander, Spain} \\
\r 7  {\eightit Enrico Fermi Institute, University of Chicago, Chicago, 
Illinois 60637} \\
\r 8  {\eightit Joint Institute for Nuclear Research, RU-141980 Dubna, Russia}
\\
\r 9  {\eightit Duke University, Durham, North Carolina  27708} \\
\r {10}  {\eightit Fermi National Accelerator Laboratory, Batavia, Illinois 
60510} \\
\r {11} {\eightit University of Florida, Gainesville, Florida  32611} \\
\r {12} {\eightit Laboratori Nazionali di Frascati, Istituto Nazionale di Fisica
               Nucleare, I-00044 Frascati, Italy} \\
\r {13} {\eightit University of Geneva, CH-1211 Geneva 4, Switzerland} \\
\r {14} {\eightit Harvard University, Cambridge, Massachusetts 02138} \\
\r {15} {\eightit Hiroshima University, Higashi-Hiroshima 724, Japan} \\
\r {16} {\eightit University of Illinois, Urbana, Illinois 61801} \\
\r {17} {\eightit The Johns Hopkins University, Baltimore, Maryland 21218} \\
\r {18} {\eightit Institut f\"{u}r Experimentelle Kernphysik, 
Universit\"{a}t Karlsruhe, 76128 Karlsruhe, Germany} \\
\r {19} {\eightit Korean Hadron Collider Laboratory: Kyungpook National
University, Taegu 702-701; Seoul National University, Seoul 151-742; and
SungKyunKwan University, Suwon 440-746; Korea} \\
\r {20} {\eightit High Energy Accelerator Research Organization (KEK), Tsukuba, 
Ibaraki 305, Japan} \\
\r {21} {\eightit Ernest Orlando Lawrence Berkeley National Laboratory, 
Berkeley, California 94720} \\
\r {22} {\eightit Massachusetts Institute of Technology, Cambridge,
Massachusetts  02139} \\   
\r {23} {\eightit Institute of Particle Physics: McGill University, Montreal 
H3A 2T8; and University of Toronto, Toronto M5S 1A7; Canada} \\
\r {24} {\eightit University of Michigan, Ann Arbor, Michigan 48109} \\
\r {25} {\eightit Michigan State University, East Lansing, Michigan  48824} \\
\r {26} {\eightit University of New Mexico, Albuquerque, New Mexico 87131} \\
\r {27} {\eightit The Ohio State University, Columbus, Ohio  43210} \\
\r {28} {\eightit Osaka City University, Osaka 588, Japan} \\
\r {29} {\eightit University of Oxford, Oxford OX1 3RH, United Kingdom} \\
\r {30} {\eightit Universita di Padova, Istituto Nazionale di Fisica 
          Nucleare, Sezione di Padova, I-35131 Padova, Italy} \\
\r {31} {\eightit University of Pennsylvania, Philadelphia, 
        Pennsylvania 19104} \\   
\r {32} {\eightit Istituto Nazionale di Fisica Nucleare, University and Scuola
               Normale Superiore of Pisa, I-56100 Pisa, Italy} \\
\r {33} {\eightit University of Pittsburgh, Pittsburgh, Pennsylvania 15260} \\
\r {34} {\eightit Purdue University, West Lafayette, Indiana 47907} \\
\r {35} {\eightit University of Rochester, Rochester, New York 14627} \\
\r {36} {\eightit Rockefeller University, New York, New York 10021} \\
\r {37} {\eightit Rutgers University, Piscataway, New Jersey 08855} \\
\r {38} {\eightit Texas A\&M University, College Station, Texas 77843} \\
\r {39} {\eightit Texas Tech University, Lubbock, Texas 79409} \\
\r {40} {\eightit Istituto Nazionale di Fisica Nucleare, University of Trieste/
Udine, Italy} \\
\r {41} {\eightit University of Tsukuba, Tsukuba, Ibaraki 305, Japan} \\
\r {42} {\eightit Tufts University, Medford, Massachusetts 02155} \\
\r {43} {\eightit Waseda University, Tokyo 169, Japan} \\
\r {44} {\eightit University of Wisconsin, Madison, Wisconsin 53706} \\
\r {45} {\eightit Yale University, New Haven, Connecticut 06520} \\
\end{center}
%
\newpage
\begin{abstract}\baselineskip=24pt

We have reconstructed the radiative decays 
$\chi_{b}(1P)~\rightarrow~\Upsilon(1S)~\gamma $ and 
$\chi_{b}(2P)~\rightarrow~\Upsilon(1S)~\gamma $
in $p \overline{p}$ collisions at $\sqrt{s} = 1.8$ TeV, and measured 
the fraction of $\Upsilon(1S)$ mesons that originate from these decays.
For $\Upsilon(1S)$ mesons with $p^{\Upsilon}_{T}>8.0$~GeV/$c$,
the fractions that come from $\chi_{b}(1P)$ and $\chi_{b}(2P)$ decays are 
$(27.1\pm6.9(stat)\pm4.4(sys))\%$ and $(10.5\pm4.4(stat)\pm1.4(sys))\%$,
respectively. We have derived the fraction of directly produced
$\Upsilon(1S)$ mesons to be $(50.9\pm8.2(stat)\pm9.0(sys))\%$.
                                        
\end{abstract}

\renewcommand{\baselinestretch}{2.0}
\vspace{0.5cm}

The large discrepancies between the charmonium production cross sections 
measured by the Collider Detector at Fermilab (CDF)~\cite{Cdfprl} and the 
predictions of the Color Singlet Model (CSM) can be explained in a
theoretical framework based on non-relativistic QCD~\cite{Octet}.
In this model, originally developed to describe rigorously the decay of 
heavy quarkonium states, the production process is factorized into 
short distance cross sections to produce the heavy quark pair, 
and long distance matrix elements, describing their binding into the 
quarkonium state. 
These matrix elements must be determined from experimental data but are 
assumed to be independent of the reaction and can be used to predict 
other processes. 
A consequence of this approach, when applied to charmonium production
in $p \overline{p}$ collisions, is the realization that $c\bar{c}$ pairs, 
produced at short distance in a color-octet state, are responsible for 
the bulk of the cross section.
In the bottomonium sector CDF has measured the inclusive production
cross section of $\Upsilon(1S)$, $\Upsilon(2S)$ and $\Upsilon(3S)$.
The prediction of CSM underestimates the measured rate, although by
a smaller amount than found for charmonium~\cite{Upsprl}.
Color-octet contributions can account for the discrepancies, but data 
on the inclusive $\Upsilon$ cross section alone is not enough to extract
the matrix elements without assumptions~\cite{Cho}. 
In order to do this one needs to separate experimentally $\Upsilon$'s 
produced directly from those arising from the decays of heavier mesons.

In this letter we report a study of the reaction 
$p\bar{p}\:\rightarrow\:\chi_b\:X $, 
$\chi_b\:\rightarrow\:\Upsilon(1S)\:\gamma $,
$\Upsilon(1S)\:\rightarrow\:\mu^+\mu^-$ at \mbox{$\sqrt{s}=1.8$~TeV} using CDF.
This analysis, based on approximately 90~pb$^{-1}$ of data collected 
during the 1994-1995 collider run, describes the first observation of 
$\chi_b$ mesons at a hadron collider.
Since the branching fractions for $\chi_b$ decays into other modes containing 
an $\Upsilon(1S)$ are expected to be small, this study allows us to measure 
the contribution of $\chi_b$ decays to $\Upsilon(1S)$ production.
Even though $\Upsilon$ mesons can be reconstructed at CDF throughout the low 
$p^{\Upsilon}_{T}$ region, we perform this measurement only for 
$p^{\Upsilon}_{T}>8.0$~GeV/$c$ because at lower $p^{\Upsilon}_{T}$ 
the photon emitted in the radiative $\chi_b$ decay is not energetic enough 
to be detected efficiently.
In this analysis we do not study transitions of $\chi_b$ mesons to 
$\Upsilon(2S)$ because photons from this decay have even lower energy.

The CDF detector has been described in detail elsewhere~\cite{cdf}.
The events used in this analysis were collected with a 
three-level trigger system which selects events consistent with 
the presence of two muons.
The first level required that two candidates be observed in
the muon chambers. 
The second level required that two or more charged particle tracks, 
partially reconstructed in the central tracking chamber (CTC) using a
fast processor, matched within $15^\circ$\ in $\phi$ (the azimuthal angle) 
the muon candidates.
The third level required better precision on the azimuthal matching and 
required the dimuon invariant mass to be between 8.5 and 11.4~GeV/$c^2$.

\begin{figure}
\hfil
\epsfxsize=6.0in
\epsffile{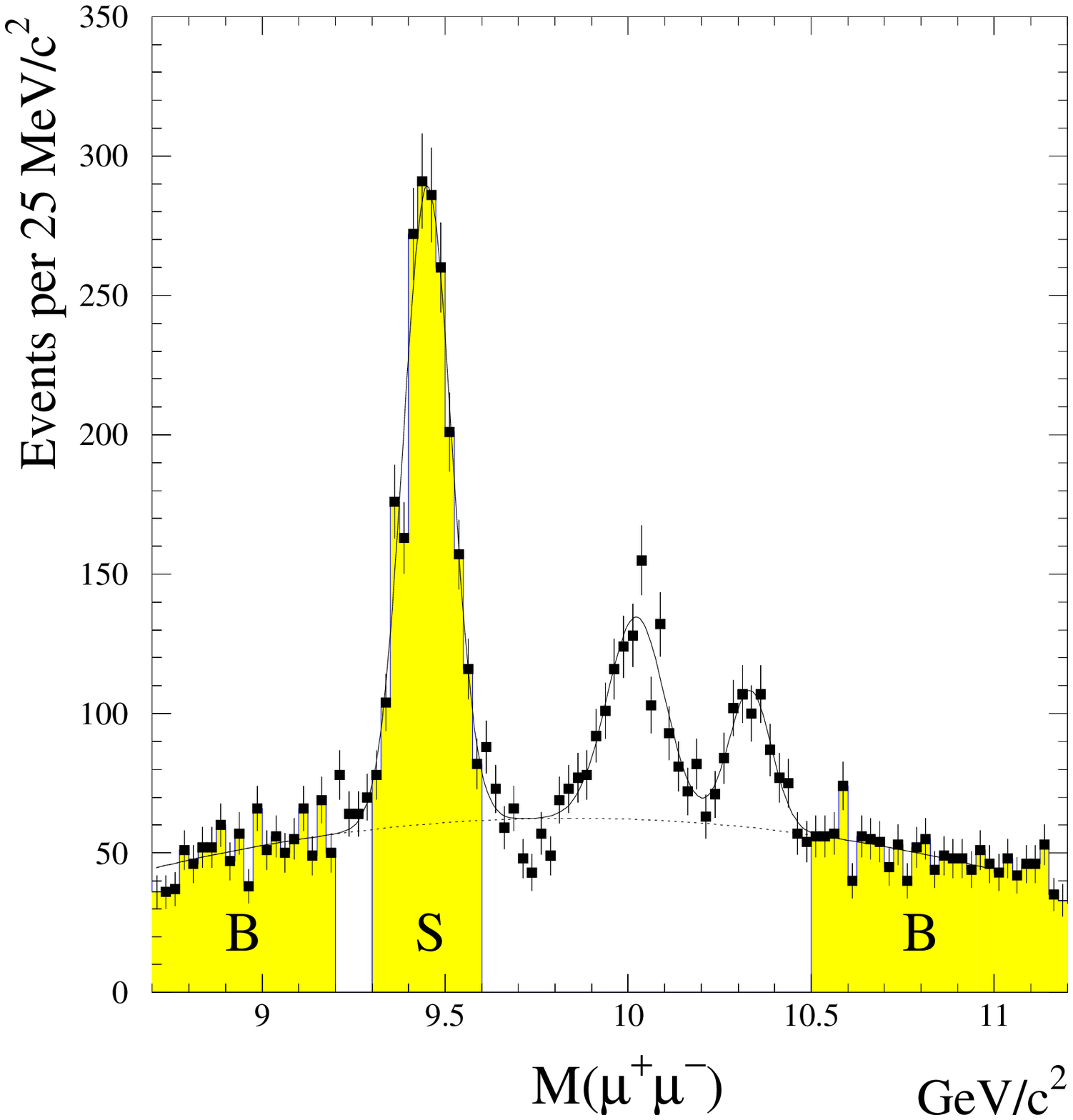}
\hfil
\caption{{\label{fig:ups}}}
\end{figure}

To identify $\Upsilon$'s we select pairs of oppositely charged muon 
candidates with $p_T>2.0$~GeV/$c$, and we require the pair to have 
$p_T(\mu^+\mu^-)>8.0$~GeV/$c$. Since $\Upsilon$ mesons do not originate from 
long lived particles~\cite{wenzel} we constrain the muon tracks to 
originate from the primary interaction point to improve mass resolution.
Fig.~\ref{fig:ups} is the resulting dimuon invariant mass distribution 
showing three peaks corresponding to the $\Upsilon(1S)$, $\Upsilon(2S)$ 
and $\Upsilon(3S)$ resonances.
Due to the trigger and muon acceptance, the pseudorapidity of the muon 
pairs is limited to the central region, corresponding approximately 
to $|\eta(\mu^+\mu^-)|<0.7$, where $\eta= -$ln[tan($\theta/2$)] and
$\theta$ is the polar angle with respect to the beam axis. 
A muon pair is considered an $\Upsilon(1S)$ candidate if its invariant 
mass is in the signal region defined by 
9300~MeV$/c^2<M(\mu^+\mu^-)<$~9600~MeV$/c^2$; this selection yields 
a sample of 2186 events.
The number of background events in this sample is obtained by fitting 
the invariant mass distribution to a polynomial plus three Gaussians and 
integrating the function associated with the background in the signal region. 
The resulting number of $\Upsilon(1S)$ mesons is $1462\pm55$.

Photon candidates are selected by demanding a transverse energy 
deposition of at least 0.7 GeV in a cell of the central electromagnetic 
calorimeter and a signal in the fiducial volume of the proportional chambers 
(CES) which are embedded in the calorimeter at a depth of six radiation lengths.
The fiducial volume requirement ensures that the shower is fully contained
in a cell.
The location of the signal in the CES chambers and the event interaction point
determine the direction of the photon momentum; its magnitude is the 
energy deposited in the calorimeter.
We correct the photon energy for the energy lost in the material in 
front of the calorimeter based on a simulation of the detector response to 
photons. For low energy photons the average correction factor varies from
1.03 to 1.14 depending on the polar angle.
We have verified that the simulation is trustworthy by comparing the 
simulated electron response with the response of electrons from photon 
conversions found in the data. 

\begin{figure}
\hfil
\epsfxsize=6.0in
\epsffile{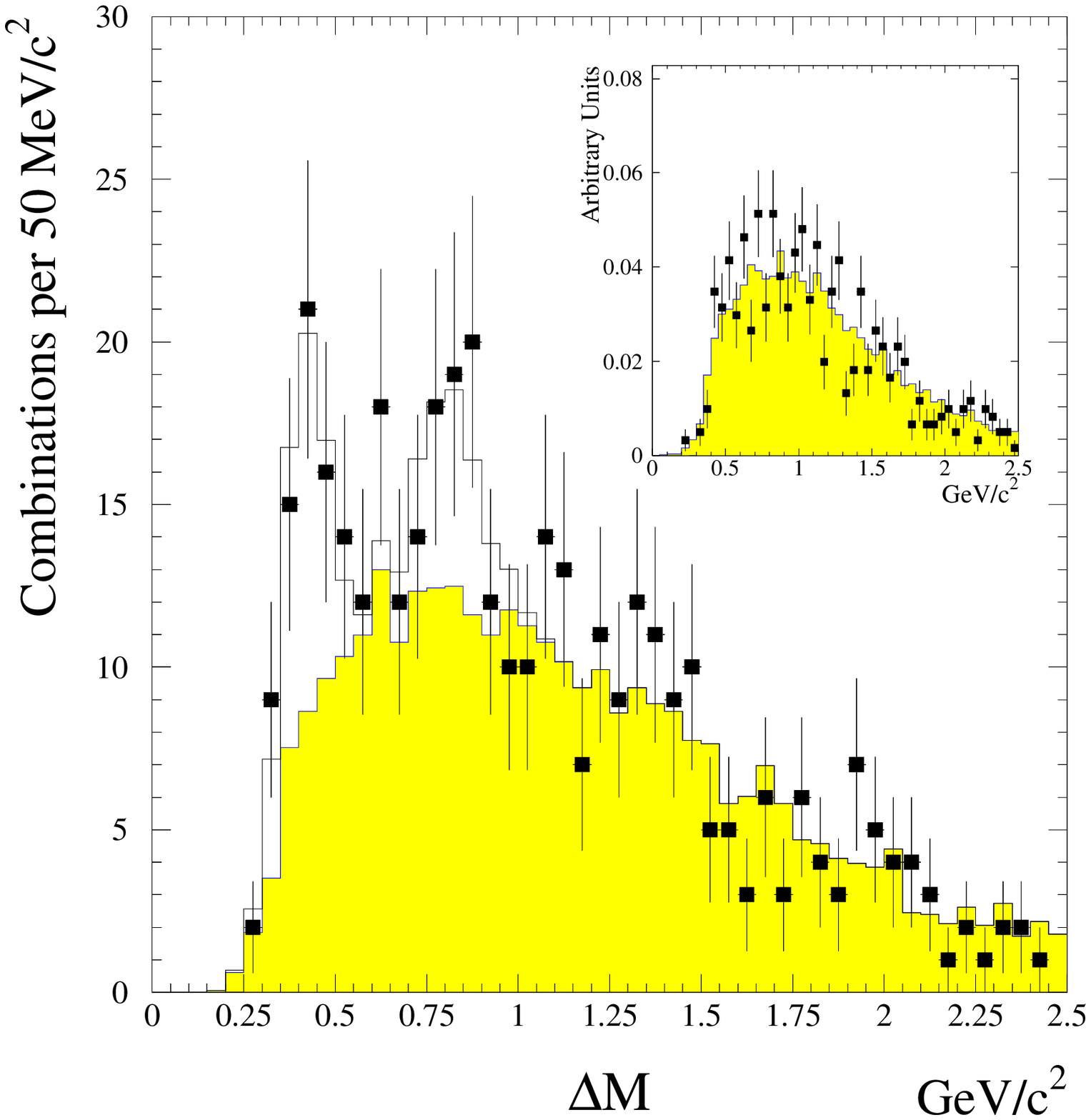}
\hfil
\caption{{\label{fig:delmas}}}
\end{figure}

To reduce the combinatorial background resulting from multiple photon
candidates per event we apply the following isolation requirements 
to the photon: 
a) no charged particle track should point to the photon cell, 
b) only one CES cluster should be associated with the cell, and
c) the total electromagnetic energy in the 8 cells neighboring the photon must
   be less than 0.5 GeV.
The $\Upsilon(1S)$ is combined with all remaining photons within the 
90 degree cone around the $\Upsilon(1S)$, and the invariant mass 
difference, $\Delta M=M(\mu^+\mu^-\gamma)-M(\mu^+\mu^-)$, is calculated. 
The $\Delta M$ distribution is shown in Fig.~\ref{fig:delmas}. 
There are two well separated signals; their masses and widths are consistent 
with expectations based on a simulation of the radiative decays of the 
$\chi_b(1P)$ and $\chi_b(2P)$ mesons. The individual angular momentum 
states of the $\chi_b$'s (J=1,2) however, cannot be resolved.

The shape of the background resulting from combinations of the $\Upsilon(1S)$
with photons unassociated with $\chi_b$ decays is obtained with a Monte Carlo 
method that uses $\Upsilon(1S)$ candidate events as input.
We consider as sources of photons:
a) decays of $\pi^0$ that are not from $\eta$ or $K^0_S$ decays, 
b) $\eta$ decays, c) $K^0_S$ decays.
These sources are simulated by replacing each charged particle in the event, 
other than the two muons, with a $\pi^0$, $\eta$ or $K^0_S$ with probabilities 
proportional to 4:2:1. These proportions follow from isospin symmetry and the 
ratios $K^{\pm}/\pi^{\pm}=0.25~$,$~\eta/\pi^0=0.5$~\cite{ratios}. 
Uncertainties in these ratios are considered as sources of systematic 
uncertainty.
The response of the detector to the photons resulting from the decay of these 
embedded neutral particles is calculated using a Monte Carlo simulation.
Applying the $\chi_b$ reconstruction to these events results in a mass 
distribution that models the shape of the background.
This model was tested by comparing the Monte Carlo distribution obtained 
using events in the mass sidebands of the $\Upsilon(1S)$ peak, with the 
corresponding distribution obtained directly from the data where there 
should be no $\chi_b$ signal.     The two distributions agree well 
as shown in the inset of Fig.~\ref{fig:delmas}.
The number of $\chi_b$ signal events is determined by fitting the data 
$\Delta M$ distribution to the sum of the background distribution, with an 
unconstrained normalization, and two Gaussian functions associated with the 
signals. The mass resolution was fixed to the value calculated by the 
simulation (60 and 93~MeV$/c^2$).
The fit results in $35.3\pm9.0$ and $28.5\pm12.0$ signal events for 
$\chi_b(1P)$ and $\chi_b(2P)$ respectively.

The fraction of $\Upsilon(1S)$ mesons originating from $\chi_b$ decays is 
calculated according 
to the equation:
\begin{eqnarray*}
{F_{\chi_b}^{\Upsilon(1S)}} = {N^{\chi_b} \over
{N^{\Upsilon} \cdot A^{\gamma}_{\Upsilon} \cdot \epsilon^{\gamma}}}
\end{eqnarray*}
where $N^{\chi_b}$ and $N^{\Upsilon}$ are the numbers of reconstructed 
$\chi_b$ and $\Upsilon(1S)$ mesons respectively, $A^{\gamma}_{\Upsilon}$ is 
the probability to reconstruct the photon once the $\Upsilon(1S)$ is found,
and $\epsilon^{\gamma}$ is the efficiency of the isolation cuts.

The photon acceptance, $A^{\gamma}_{\Upsilon}$, is the product of the 
probability that the photon is within the fiducial volume and the 
reconstruction efficiency of the fiducial photon.
The geometric acceptance is determined by using a Monte Carlo simulation
where $\chi_b$'s are generated uniformly in pseudorapidity, and with a $p_T$ 
distribution equal to the measured $\Upsilon(1S)$ spectrum~\cite{Upsprl}. 
The $\chi_b\:\rightarrow\:\Upsilon(1S)\:\gamma$ decay is generated 
with a uniform angular distribution in the $\chi_b$ rest frame.
The $\Upsilon(1S)\:\rightarrow\:\mu^+\mu^-$ decay is also generated uniformly 
in the $\Upsilon(1S)$ rest frame and the trigger simulation is applied to the 
decay muons. Uncertainties associated with the $p_T$ spectrum used for the 
production of $\chi_b$ mesons and with the unknown $\chi_b$ polarization
are considered as sources of systematic uncertainty.
The photon reconstruction efficiency is obtained from the data by
applying the photon requirements, except for the isolation cuts,
to a sample of electrons from photon conversions selected using only tracking
information. This efficiency is then corrected for the known differences in the
detector response between photons and electrons. 
The reconstruction efficiency rises from 17\% to 85\% for photon with $E_T$ 
ranging from 0.7~GeV to 1.4~GeV.
For $p^{\Upsilon}_T>8.0$~GeV/$c$, the photon acceptance is 
$0.142\pm0.004(stat)$ and $0.284\pm0.006(stat)$ for $\chi_{b}(1P)$ 
and $\chi_{b}(2P)$ respectively. The large difference is entirely due to 
the mass difference between the parent particles, resulting in different 
photon energies.

To study the effect of the isolation cuts we use a Monte Carlo method 
that uses $\Upsilon(1S)$ candidate events as input.
For each event, we generate a vector distributed according to the 
angular distribution of the photon, relative to the $\Upsilon(1S)$ momentum, 
obtained simulating the decay $\chi_b\:\rightarrow\:\Upsilon(1S)\:\gamma$.
The probability that the isolation requirements are satisfied when applied to 
the calorimeter cell intercepted by the vector gives the cut efficiency.
Since there are background events in the $\Upsilon(1S)$ signal region, 
we measure the efficiency in the signal and sideband regions and derive 
the efficiency associated with $\Upsilon(1S)$ mesons.
The resulting efficiency is $\epsilon^{\gamma}=0.627\pm0.013(stat)$ 
for $\chi_b(1P)$ and $\epsilon^{\gamma}=0.651\pm0.013(stat)$ for 
$\chi_b(2P)$; the difference is due to the different kinematic of the decays.
We assume that this efficiency, calculated from the inclusive sample 
of $\Upsilon(1S)$ events, is applicable to the subsample of interest where 
the $\Upsilon(1S)$ originates from a $\chi_b$. 
This assumption is supported by a study using samples of $J/\psi$ events. 
We calculate $\epsilon^{\gamma}$ using the inclusive sample of $J/\psi$ 
events, with the Monte Carlo method just described, and independently using 
a pure sample of $J/\psi$ from $\chi_c$ decay.
The latter is the sample of $\chi_c\:\rightarrow\:J/\psi\:\gamma$ reconstructed 
by requiring the photon to convert into an electron-positron pair. 
In this sample we measure the efficiency by applying the isolation cuts 
to the calorimeter cell which would have been hit by the photon, had it not 
converted~\cite{conv}.
This measurement yields an efficiency of $0.57\pm0.06(stat)$; the Monte Carlo 
calculation is in good agreement, yielding an efficiency of 
$0.56\pm0.01(stat)$.

The systematic uncertainty on $F_{\chi_b}^{\Upsilon(1S)}$ associated with the 
$\chi_b$ production and decay model is estimated by varying the shape  of the
$p_T$ spectrum as well as the decay angular distribution to account  for fully
polarized $\chi_b$'s; the uncertainty is $\pm13\%$ for $\chi_{b}(1P)$  and
$\pm9\%$ for $\chi_{b}(2P)$. The uncertainty in the determination of
$N^{\chi_b}$ is $\pm7\%$ for  $\chi_{b}(1P)$ and $\pm9\%$ for $\chi_{b}(2P)$.
This includes the effect of varying the $\pi^0,\eta$ and $K^0_S$ composition 
in our background model from 4:2:1 to all $\pi^0$, and a variation of $\pm2\%$ 
of the calorimeter energy scale used in the simulation. 
It also includes the effect of varying the resolution of the Gaussians used 
in the fit by $\pm6\%$, the uncertainty on the resolution. 
An uncertainty of $\pm6\%$ for $\chi_{b}(1P)$ and $\pm 3\%$ for $\chi_{b}(2P)$
is associated with the estimation of the detector response difference between 
photons and electrons. 
An additional $\pm4\%$ uncertainty arises from the
statistical and systematic uncertainties associated with $\epsilon^{\gamma}$.
We combine these uncertainties, assuming they are independent, into a total
systematic uncertainty of $\pm 16.4\%$ for $\chi_{b}(1P)$ and $\pm 13.7\%$ for
$\chi_{b}(2P)$. The fractions of $\Upsilon(1S)$ mesons, with
$p^{\Upsilon}_{T}>8.0$~GeV/$c$, that come from $\chi_{b}(1P)$ and 
$\chi_{b}(2P)$ decays are  $(27.1\pm6.9(stat)\pm4.4(sys))\%$ and
$(10.5\pm4.4(stat)\pm1.4(sys))\%$ respectively. 

To calculate the fraction of directly produced $\Upsilon(1S)$ mesons we 
must estimate the fraction of $\Upsilon(1S)$'s associated with sources other 
than $\chi_{b}(1P)$ and $\chi_{b}(2P)$.
We calculate the contribution due to 
$\Upsilon(2S),\Upsilon(3S)\:\rightarrow\:\Upsilon(1S)\:\pi\pi$ using a 
Monte Carlo simulation of these decays normalized with the $\Upsilon(2S)$ 
and $\Upsilon(3S)$ cross section measured in this experiment~\cite{Upsprl}.
We find that the fraction of $\Upsilon(1S)$'s, with 
$p^{\Upsilon}_{T}>8.0$~GeV/$c$, from $\Upsilon(2S)$ and $\Upsilon(3S)$ decays,
is $(10.7^{+7.7}_{-4.8})\%$ and $(0.8^{+0.6}_{-0.4})\%$ respectively.
An additional contribution could be associated to the yet unobserved 
$\chi_b(3P)$ mesons. These states are predicted to lie below $B\bar{B}$ 
threshold and to decay radiatively to $\Upsilon(1S)$, $\Upsilon(2S)$ 
and $\Upsilon(3S)$.
An upper limit on the fraction of $\Upsilon(1S)$'s from $\chi_b(3P)$ decays
can be calculated with the conservative assumption that all $\Upsilon(3S)$ 
mesons in our data come from $\chi_b(3P)$ decays.
To estimate the contribution to $\Upsilon(1S)$, relative to $\Upsilon(3S)$, 
we have used a theoretical calculation of the radiative decay widths
of the $\chi_b(3P)$~\cite{silverman} and the detector simulation to take 
into account the effect of the trigger and kinematical cuts.
Our estimate is that less than $6\%$ of the $\Upsilon(1S)$'s, with 
$p^{\Upsilon}_{T}>8.0$~GeV/$c$, arise from $\chi_b(3P)$ decays.
We derive the fraction of directly produced $\Upsilon(1S)$ mesons
according to the equation ${F_{dir}^{\Upsilon(1S)}} =
{1-F_{\chi_b}^{\Upsilon(1S)}-F_{\Upsilon}^{\Upsilon(1S)}}$, where 
$F_{\Upsilon}^{\Upsilon(1S)}$ is the fraction of $\Upsilon(1S)$'s
from $\Upsilon(2S)$ and $\Upsilon(3S)$.
Systematic uncertainties on $F_{dir}^{\Upsilon(1S)}$ arise from 
uncertainties on the $\Upsilon(2S)$ cross section and branching fractions.
Our upper limit on the contribution from $\chi_b(3P)$ decays is also considered 
a systematic uncertainty, and is added in quadrature to the negative error.
We find $F_{dir}^{\Upsilon(1S)} = (50.9\pm8.2(stat)\pm9.0(sys))\%$ 
for $p_T^\Upsilon>8.0$~GeV/$c$.

In conclusion, we have measured the fraction of $\Upsilon(1S)$ mesons
originating from $\chi_b$ decays and derived the fraction of 
directly produced $\Upsilon(1S)$'s.
We find that $(27.1\pm6.9(stat)\pm4.4(sys))\%$ of all $\Upsilon(1S)$ mesons 
with $p^{\Upsilon}_{T}>8.0$~GeV/$c$ come from $\chi_{b}(1P)$ decays,
$(10.5\pm4.4(stat)\pm1.4(sys))\%$ from $\chi_{b}(2P)$ decays, and 
$(50.9\pm8.2(stat)\pm9.0(sys))\%$ are directly produced. 
A calculation based on the Color Singlet Model~\cite{adam} predicts
a contribution of about $41\%$ and $13\%$ from $\chi_{b}(1P)$ and 
$\chi_{b}(2P)$ decays respectively.
This measurement will allow the determination of the matrix elements 
associated with the production of $\chi_{b}(1P)$, $\chi_{b}(2P)$ and 
$\Upsilon(1S)$ mesons, thus providing information on color-octet 
contributions in bottomonium production.

We thank the Fermilab staff and the technical staffs of the
participating institutions for their vital contributions.  This work was
supported by the U.S. Department of Energy and National Science Foundation;
the Italian Istituto Nazionale di Fisica Nucleare; the Ministry of Education,
Science and Culture of Japan; the Natural Sciences and Engineering Research
Council of Canada; the National Science Council of the Republic of China;
and the A. P. Sloan Foundation.

\renewcommand{\baselinestretch}{1.0}

\newpage
\centerline {\large\bf Figure Captions}
\begin{list}{}
\item{FIG. 1.}
The invariant mass distribution of muon pairs after the selection
described in the text. Region S is the $\Upsilon(1S)$ signal region,
region B defines the $\Upsilon(1S)$ sidebands.
The solid line is the function used to fit the data, the dotted
line is the function used to calculate the number of background events
in the signal region.
\item{FIG. 2.}
The mass difference distribution, 
$\Delta M=M(\mu^+\mu^-\gamma)-M(\mu^+\mu^-)$,
after the selection described in the text. 
The points represent the data. The shaded histogram is the background 
shape predicted by the Monte Carlo calculation. 
The solid line is the fit of the data to two Gaussian functions plus the 
background histogram. 
The inset shows the comparison between the $\Delta M$ distribution
for dimuons in the $\Upsilon(1S)$ sidebands (region B of Fig.1), 
and the corresponding one predicted by the Monte Carlo calculation; 
the two distributions are normalized to equal area and the vertical 
scale is arbitrary. The size of the bin is the same in both figures.
\end{list}

\newpage

\newpage

\end{document}